\documentclass{aastex}                    
\usepackage{epsfig}
\usepackage{psfig}

\newcommand{\thoric}{$\theta^1$~Ori~C}
\newcommand{\iue}{{\it IUE}}
\newcommand{\kms}{km\,s$^{-1}$}
\newcommand{\teff}{$T_{\rm eff}$}

\newcommand{\rstar}{$R_\star$}
\newcommand{\tnd}{\tablenotemark{d}}

\slugcomment{Version of \today}
\shortauthors{Smith \& Fullerton}
\shorttitle{Magnetic Wind of \thoric}

\begin{document}

\title{A Revised Geometry for the Magnetic Wind of $\theta^1$~Orionis C }

\author{Myron A. Smith}
\affil{Catholic University of America,\\
        3700 San Martin Dr., 
	Baltimore, MD 21218;           \\
	msmith@stsci.edu } 

\and

\author{Alex W. Fullerton\altaffilmark{1}}
\affil{Dept. of Physics \& Astronomy, 
       University of Victoria,                 \\ 
       P.O.\ Box 3055,  
       Victoria, BC, V8W 3P6, Canada.}
\altaffiltext{1}{Postal Address: Dept. of Physics \& Astronomy, 
                 The Johns Hopkins University, 
                 3400 N. Charles St.,
                 Baltimore, MD 21218; awf@pha.jhu.edu }

\begin{abstract}

The star $\theta^1$\,Orionis\,C (O6--7\,V) is often cited as a hot analog of 
Bp variables because its optical and UV line and X-ray continuum fluxes 
modulate over the magnetic/rotational period.
In this circumstance, one expects emission and absorption components of the 
UV resonance lines to vary as a flattened magnetosphere co-rotates with the
star. 
In this paper we re-examine the detailed velocity behavior of several 
strong UV lines.
Whereas past work has focused on variations of the full profiles, we 
find that the blue and red wings of the \ion{C}{4} and \ion{N}{5}
resonance lines exhibit anticorrelated modulations.
These appear as absorption excesses at large blueshifts, and
flux elevations at moderate redshifts at the edge-on phase $\phi$=0.5.
No rest-frame absorption features, which are the typical signatures 
of cool, static disks surrounding Bp stars, can be detected at any phase.

We suggest that this behavior is caused by two geometrically distinct
components of the wind, which are defined by the relationship between
the extent of a magnetic loop and the local Alfven radius.
Streams on field lines opening outside this radius are first channeled 
toward the magnetic equator, but after reaching the Alfven radius they are 
forced outward by radiative forces, eventually to become an expanding radial 
outflow.
This wind component causes blueshifted absorption as the co-rotating 
magnetic equatorial plane crosses the observer's line of sight ($\phi$= 0.5).
The geometry of the inner component requires a more complicated interpretation. 
Wind streams first follow closed loops and collide at the magnetic equator
with counterpart streams from the opposite pole.
There they coalesce and fall back to the star along their original field lines.
The high temperatures in these falling condensations cause the 
redshifted emission.
The rapid circulation of these flows is likely the reason for the absence of 
signatures of a cool disk (e.g., zero-velocity absorptions at $\phi$ $\sim$ 0.5)
in the strong UV lines. 

\end{abstract}


\section{Introduction}

The discovery by Donati et al. (2002) of a magnetic field in the periodic 
variable {$\theta^1$\,Orionis\,C} (HD\,37022; O6--7 V: Morgan \& Keenan 1973; 
O7: Stahl et al. 1996) solidified the expectation that it is a hot analog 
of the magnetic B-type variables, the so called He-anomalous Bp stars. 
In contrast to the Bp stars, the modulation of soft X-ray flux in {\thoric}
reveals the presence of a hot, co-rotating magnetosphere (Gagn\'e et al. 1997). 
The magnetic and X-ray periods are presumed to correspond to the 
rotational period of 15.422 days, as determined from ultraviolet and 
optical line variations (Stahl et al. 1993, Stahl et al. 1996, 
Reiners et al. 2000). 
Nonetheless, questions have lingered as to how to reconcile
the phased variations of strong optical and UV lines with the geometry 
of the Bp magnetosphere model (Donati et al. 2002).

The magnetic Bp variables can be classified as a subgroup of a larger class of 
early-type stars that includes at least two pulsating $\beta$\,Cephei stars 
(Henrichs et al. 1998), the ``sn" stars\footnote{The sn stars are a subgroup 
of the He-weak Bp stars. They are so named because of the broad or 
``nebulous" appearance of the helium lines in their spectra as well as
the sharp, central components in the hydrogen and \ion{C}{2}~$\lambda$4267 
lines.} recently examined by Shore et al. (2004), and 
other related objects (e.g., Smith \& Groote 2001; hereafter SG01).
According to the usual picture, winds in magnetic Bp stars are prevented 
from escaping from the magnetic poles, but are channeled outwards along 
dipolar field loops. 
The wind is constrained to flow along these loops until it encounters streams 
from the opposite pole.
When the two streams collide, they shock, cool, and either circulate back to 
the star or leak out through the outer disk edge where they are lost to the 
system. 
The initial formulation of this picture by Shore and collaborators 
(e.g., Shore \& Brown 1990; hereafter SB90) envisaged polar 
outflows (``jets") and a cool, toroidal disk. 

Subsequent magnetostatic modeling of the wind flows of Bp stars 
by Babel \& Montmerle (1997) has dealt with the post-shock cooling of 
the gas in the magnetospheric disk as the principal site of the UV 
and H$\alpha$ line variations. 
In this formulation, rotation is so small as to be negligible.
The frozen-in magnetosphere then resembles a wobbly inner tube that
co-rotates with the Bp star. 
If the disk is extended enough in radius, it can contribute substantial 
line emission when it is viewed face-on. 
Moreover, disk regions passing in front of the star generally have a lower 
temperature than the photosphere and therefore remove additional flux from 
photospheric absorption lines.
In the case of resonance lines, scattering of line flux has the same effect.
Thus, over the cycle various viewing aspects of the wind/disk complex can 
produce a variety of emissions and absorptions in the UV resonance 
lines of \ion{Si}{4}, \ion{C}{4}, and \ion{N}{5}.
In addition, modulations of H$\alpha$ emission (Short \& Bolton 1994) and 
high-level Balmer line absorptions (Groote \& Hunger 1976) provide excellent 
diagnostics for well-developed magnetospheres, such as the one
surrounding the prototypical He-strong star $\sigma$\,Orionis\,E. 
Along the hot edge of the classical Bp domain, dense winds 
insure that the wind particles do not fractionate chemically, with the result 
that the surface abundances of these stars remain normal.

The study of the wind geometry of {\thoric} is compelling because it is the 
only O star currently known to possess a magnetic field
(Henrichs 2005). 
The presence of a strong field in a dense, high-velocity wind make its 
circumstellar environment a unique laboratory to study high-energy 
magnetohydrodynamic plasma (Babel \& Montmerle 1997).  
The interaction of various components of the wind produces the excitation of 
hot gas visible in X-ray spectra (Gagn\'e et al. 2005) and potentially in the 
far-UV.  Also, despite the inhibition caused by the strong magnetic field, 
the mass-loss rate of {\thoric} is several hundred times larger
than then mass-loss rates typical of hot Bp stars (Donati et al. 2002).
The strong stellar wind, together with the star's slow rotation rate, permit
spectral features to be Doppler shifted outside the photospheric line 
profile, which provides an opportunity to map the geometry of the wind over
a wide range of stellar distances and azimuths.

\section{A Sketch of a New Magnetic Geometry}\label{sktch}

In this study we adopt the values for the rotational and magnetic 
tilt angles $i$ and $\beta$ from the analysis of the longitudinal 
magnetic field curves by Donati et al. (2002): both parameters are close 
to 45{\degr}. 
These values are the same as those adopted by Stahl et al. (1996; hereafter 
S96) and nearly the same to those of Reiners et al. (2000; hereafter R00). 
However, as noted by Donati et al. (2002), the mapping between line
variations and magnetic phase adopted by S96 implies that the magnetic
pole is viewed face-on at $\phi$ = 0.5, which is difficult to reconcile
with the observed H$\alpha$ absorption at this phase.
Instead, the magnetic-field measurements of Donati et al. (2002) show
conclusively that phase $\phi$ = 0.0 corresponds to the passage of
the north magnetic pole across the projected center of the star.
At this phase, material in the equatorial plane is seen ``face-on" by a distant 
observer.
At $\phi$ = 0.5, the north magnetic pole has rotated to a point on the star's 
``upper" limb and the south magnetic pole makes its only brief appearance 
at the star's ``lower" limb. 
At this time the observer sees the post-shock magnetosphere (i.e., the material
near the magnetic equator) edge-on. 
If the Bp paradigm of SB90 and SG01 were applicable, maximum H$\alpha$ emission
and excess absorption in UV resonance lines should be observed at $\phi$ = 0 
and 0.5, respectively. As we will see below, this is not the case. 
Furthermore, Donati et al. (2002) were unable to explain the 50~{\kms} 
blueshift in the centroid of the H$\alpha$ emission component in the context 
of the Bp picture.

The configuration described in this paper, shown as a sketch in 
Figure\,\ref{cartoon}, is much different from this standard Bp paradigm. 
Our geometry attributes the blueshifted absorption in the \ion{C}{4} resonance 
lines at $\phi$ = 0.5 to a radially outflowing, high-velocity wind. 
The point marked ``X" designates schematically the Alfven radius at the 
magnetic equator. 
For a very slowly rotating star this is where a bifurcation occurs in wind 
stream trajectories (e.g., ud-Doula \& Owocki 2002). 
All streams sketched in Fig.\,\ref{cartoon} are diverted toward the magnetic 
equator because the particles are constrained at first to flow along field 
lines by the dominant magnetic forces.  
However, wind streams originating from high magnetic latitudes follow less 
confined loops that eventually extend beyond the Alfven radius. 
As these loops traverse the Alfven radius, radiation pressure (which falls off 
less quickly with distance from the star) overwhelms the magnetic forces and 
drags the field radially outward. 
Thus, these streams evolve to become a concentrated outflow confined to 
low magnetic latitudes. 
In contrast, streams in the inner wind zone that originate from intermediate 
magnetic latitudes on the star remain confined and contribute or remove line 
flux only when they collide with one another, shock, and return to the star
(\S~\ref{infall}).  
The fate of this wind component contrasts strongly with the formation of the 
quasi-static, corotating disk typical of Bp stars, and is the subject of much 
of the rest of this paper.

\section{The Morphology of Spectral Features}

\subsection{Line profile morphology according to previous studies}

In their seminal paper, S96 demonstrated that the emissions in H$\alpha$, 
\ion{He}{2}~$\lambda$4686, and the absorptions of the \ion{C}{4} and \ion{Si}{4}
 resonance doublets of {\thoric} are regularly modulated over a period of
15.422 days. 
This value is assumed to be the stellar rotational period. 
Phases in magnetic Bp variables are traditionally reckoned from the time of 
magnetic maximum, i.e., from  the transit of a magnetic pole across the 
observer's line-of-sight.
Typically for these stars, if magnetic ephemerides are not available, the 
zero point is taken as the epoch of H$\alpha$ emission maximum (equivalent 
width minimum). 
Because the central axis of the magnetosphere coincides with the magnetic 
plane, these zero points coincide.

Using this zero point and period, S96 and R00 measured the equivalent-width 
variations with phase for several strong UV resonance and optical excited 
lines.
The variations exhibit a maximum at $\phi$ = 0, and both groups of
authors interpreted this as excess absorptions in the line cores. 
The variations in the UV resonance, H$\alpha$, H$\beta$ (G. Wade, private
communication), 
and \ion{He}{2}~$\lambda$4686 lines are also highly nonsinusoidal. 
As measured across the entire profile, the phased equivalent width variations 
have an ``M"-shape centered at $\phi$ = 0.5. 
The second maximum of the ``M" is centered at $\phi$ $\approx$ 0.6--0.7
and is generally stronger than the first prong at $\phi$ = 0.4. 
In the case of the H$\alpha$ line, the local minimum at $\phi$ = 0.5
arises at least partly from the brief appearance of a red emission 
component (Stahl et al. 1993).

\subsection{A detailed look at the \ion{C}{4} and \ion{N}{5} resonance lines}

To reinvestigate the variations exhibited by the UV lines, particularly
the \ion{C}{4} and \ion{N}{5} resonance doublets, we retrieved the
22 high-dispersion {\iue} spectra of {\thoric} obtained through the
large aperture with the SWP camera from the Multi-Mission Archive at
Space Telescope\footnote{MAST is supported for the archiving of UV/optical 
data obtained by NASA satellites other than HST by the NASA Office of 
Space Science through grant NAG5-7584.}.
The basic features of the variations are shown in Figure~\ref{grysc}, 
which llustrates the fluctuations in \ion{C}{4} as a function phase in 
in the form of a grayscale (``dynamic") spectrum.
The format of Fig.~\ref{grysc} is similar to Figure~10 of S96,
except that the mean profile has been subtracted from individual
spectra in the time series rather than the profile near $\phi = 0$,
and the difference spectra illustrated in the image have been 
smoothed over {1~\AA} intervals.

Fig.~\ref{grysc} shows that the  variations in the \ion{C}{4}
resonance doublet are dominated by enhanced absorption (darker shades)
in the high-velocity region of the unusually shallow P~Cygni absorption
trough. However, weaker variations are also present throughout the absorption 
trough and even extend into the redshifted part of the profile. The red-wing
variations of the $\lambda$1548 line are confused by blending between 
the doublet components, but the corresponding variations of the 
$\lambda$1550 can be seen clearly in window ``e" in the figure. 
As discussed below, the weaker variations exhibit different behavior as 
a function of phase than the high-velocity absorption excess.

\subsubsection{Maximum and minimum absorption profiles}

The \ion{C}{4} equivalent width-phase curves (e.g., Figure\,8 in S96) 
show some interesting features, and we have already
mentioned the ``M-shaped morphology. These features, particularly those 
pertaining to the different dependences of fluxes across the line profile, 
can be investigated further by comparing the individual spectra at strategic 
phases. A prime example of this is the different manner in which the 
fluxes across the line profiles contribute to the equivalent width of 
the full profile. Since the maximum swing in equivalent widths occurs 
over the comparatively narrow phase range between 0.6 and 0.9, it is 
illuminating to compare the three available spectra at these phases. 
These spectra are shown in Figure~\ref{c4mnmx}. The large flux 
variations at wavelengths shorter than {1543~\AA} are the most obvious
differences. However, we also note 
that the fluxes in the blue and red regions of 
the doublet vary in opposite senses. Notice that while profiles at 
$\phi$ = 0.5\,--\,0.6 exhibit increased blue absorption relative to the 
profiles observed near $\phi$ =0.0, the red-wing flux of the \ion{C}{4} 
$\lambda$1550 line (the region labeled ``e'') is high relative to the 
wing in the observation at $\phi$= 0.94. Put another way, the flux 
of the red wing of $\lambda$1550 is high when the fluxes of the (blended) 
blue wing of the doublet is low, and vice versa, over the rotational cycle. 
The blue wing variations overwhelms the anticorrelated red-wing ones
for the \ion{C}{4} lines, which is why a depiction of the variation of
the equivalent width extracted from the full profile conceals the red
wing variation. 

Figure~\ref{n5mnmx} likewise shows observations of the 
\ion{N}{5}~$\lambda\lambda$1238, 1242 at the same phases as the previous 
figure. The blue-wing flux variations are again present, but in this case 
they are much smaller than for the \ion{C}{4} doublet. In fact, 
it is difficult to see a variation at all between phases 0.6 and 0.9. 
However, the red-wing variation, (unblended in the two \ion{N}{5} components), 
as indicated also by the smoothed difference 
spectrum at the bottom of the figure, is clearly visible.  

  Another interesting feature of Figures\,\ref{c4mnmx} and \ref{n5mnmx}
is that when allowance is made for the different local continuum levels, 
the amplitude of the red-wing variations is virtually the same for 
both components of the \ion{C}{4} and \ion{N}{5} doublets.
Although there are additional complications due to blending, the
implication is that the variable component is formed in an optically
thick medium.

\subsubsection{Variation of the lines over the rotation cycle}

 The phase-resolved fluctuations in different regions of the \ion{C}{4} 
resonance doublet just discussed can be compared by plotting 
pseudo-equivalent widths extracted from different wavelength intervals.
To provide quantitative measures of the line absorptions, we computed 
``equivalent width indices," which are summed ``net" fluxes in a defined 
series of windows that span the line.
We used the net fluxes to avoid the errors introduced by the ripple
correction for high-dispersion {\iue} data processed with ``NEWSIPS."
The equivalent width index in a given window was computed as the ratio
of the total flux in it to the total flux in the particular
order of the echelle spectrum.
Like genuine equivalent widths, larger, positive values of these indices
indicate greater absorption.
They can be measured with precision since they are independent of continuum 
placement errors, but they cannot be converted to true equivalent widths 
in Angstroms. 
Error bars were determined from the median point-to-point fluctuations in
the values of the indices.
This criterion likely overestimates the uncertainties because of the 
slow variations of the index through the cycle.

When summed over the entire profile, our plots of the \ion{C}{4} indices 
mimic the equivalent-width curves presented by S96. However, the behavior 
in individual wavelength bins distributed across the weak P~Cygni profile 
is markedly different.
For example, the shortest wavelengths are dominated by a single,
large-amplitude variation with maximum absorption near $\phi$ = 0.5.
In contrast, the indices formed from bins that sample the core and 
red wing reflect systematic, small variations in the red wing, 
as suggested by Fig.~\ref{c4mnmx}.
These red-wing variations are illustrated in Figure~\ref{c4n5phas}
for \ion{C}{4} and \ion{N}{5}. From this figure it is apparent that a 
sinusoidal variation (shown as a dashed line for reference) is not a 
good fit to the data. Fig.~\ref{c4n5phas} also shows that the red-wing 
variations of the \ion{N}{5}~$\lambda$1242 profile (0\,--\,+250~\kms) 
track the red-wing variations of the \ion{C}{4} doublet.\footnote{
Analogous red-wing variations are not present in the \ion{Si}{4} 
resonance doublet, which is primarily a photospheric feature for O7 
dwarfs.} 

We also find that a simple blue/red-profile dichotomy is an 
inadequate description of the variations of different regions of the 
\ion{C}{4} doublet with phase.
In Figure\,\ref{c4wng} we represent fluxes in four additional 
1-Angstrom bins at increasingly blueshifted portions of the absorption trough.
At the bottom of the figure we replot the curve from the red-wing 
extraction of 1550\,\AA\ in shown in Fig.\,\ref{c4n5phas}. 
The``M"-pattern is present in curves {\it b} and {\it c}, but in 
curve {\it a,} representing the most negative velocities, the two prongs 
have merged together. 
The same behavior is evident in the dynamic spectrum (Fig.~\ref{grysc}).

\subsubsection{The high- and intermediate-velocity outflows}

According to the revised geometry of the wind flow, 
spectroscopic observers in the right-hand section of Fig.~\ref{cartoon} 
will see the effects on the \ion{C}{4} line profiles of a collimated, 
high-velocity wind rotating into their line of sight as phase $\phi$ = 0.5 
approaches. Thus, the equivalent width indices extracted from the far-blue 
wing exhibit maximum absorption strength at this phase.
The ``M"-shaped curves for the near-blue wing and 
smaller velocities can also be explained by this same flow geometry. 
Between phases $\phi$ = 0.3--0.4 and 0.6--0.7, the absorptions occur
at positions in the line that correspond approximately to the wind terminal 
velocity multiplied by the cosine of the angle between the line-of-sight to 
the center of the star and the bisector line in the equatorial plane. 
Since the column responsible for the high velocity absorption at 
$\phi$ = 0.5 is shifted to intermediate velocities just before and after this
phase, a distant observer sees the maximum absorption at these times.
The wind flows seen against the rest of the stellar disk at these times 
are changing their direction and are also accelerating rapidly. 
Consequently, the absorptions these flows impose on the line profile 
are distributed over a broader velocity range and thus have a smaller 
influence on the equivalent widths extracted from any one narrow
wavelength window in the absorption trough.

\subsection{Phase behavior of excited lines}\label{optcl}

\subsubsection{The wind lines} 

In Table\,1 we summarize the relevant properties of a selection of
optical and UV lines of $\theta$$^1$\,Ori\,C reported in the recent
literature.  All of these lines show a phase modulation.
We include the phases at which absorption equivalent widths of 
various lines attain their maximum absorption values. 
Those lines for which the ``M"-morphology is most prominent are denoted by 
phases ``0.4\,--\,0.6." 
We also show two entries for the \ion{C}{4}~$\lambda\lambda$1548,1550 doublet. 
Following S96, the first of these refers to the equivalent width across the 
full profiles, while the second describes the results for the red half of 
the profile only.
A clear division can be made between the lines predominantly formed 
in the wind (i.e., the UV resonance lines, H$\alpha$, the emission in H$\beta$,
and \ion{He}{2}~$\lambda$4686) and the lines largely formed in the stellar 
photosphere (e.g., the excited transitions of \ion{C}{4} and \ion{O}{3}).

As already noted, H$\alpha$ and \ion{He}{2}~$\lambda$4686 exhibit complicated
emissions. 
At $\phi$ = 0.0, a strong emission appears in the blue wing, and subtle 
variations interpreted by Stahl et al. (1993) as emission appear at 
$\phi$ =  0.5 to the red of line center. 
According to the sketch in Figure\,\ref{cartoon}, blueshifted emissions might 
be observed as the magnetic pole crosses the projected center of the star at 
$\phi$ = 0; this possibility is addressed below.
For now we note that the appearance of strong H$\alpha$ emission (slightly 
blueshifted or otherwise, and discounting a nebular contribution) is unusual 
in the spectra of mid-O dwarfs.
Its presence in spectra of {\thoric} is certainly attributable to 
the unusual conditions in the magnetically-channeled wind.

\subsubsection{Other excited lines: optical}\label{optred}

S96 and R00 illustrated the modulations of several excited lines of
\ion{He}{1}, \ion{He}{2}, \ion{C}{4}, and \ion{O}{3}.
Depending on the data quality, the absorption curves of these lines
exhibit either a broad plateau centered at $\phi$ = 0.5, or a shallow local 
``M"-minimum at this phase. 
Although these authors constructed these curves by measuring the whole 
profile, differentiation with velocity can be found in the 
dynamic spectra of \ion{O}{3}~$\lambda$5592 and \ion{C}{4}~$\lambda$5812 
presented in these papers. 
These depictions show the blue-to-red migration of an apparent absorption, 
with line center crossing at $\phi \approx 0.75$. 
The traditional interpretations for the origins of migrating subfeatures 
in line profiles of hot stars include surface spots due to chemical 
inhomogeneities, photospheric velocity fields due to nonradial pulsations,
or circumstellar clouds forced into co-rotation. 
The first of these alternatives is improbable in an O star having a chemically 
homogeneous wind. 
The second is problematical for several reasons, but in particular 
because the migrating features are present during only half the cycle. 
We will return presently to the third alternative, ``co-rotating clouds,"
in the modified form of dense, infalling material. 

\subsubsection{Other excited lines: \ion{O}{5} $\lambda$1371 and \ion{N}{4} 
               $\lambda$1718}\label{uvexc}

S96 noted that the strength of the excited \ion{O}{5}~$\lambda$1371 line 
exhibits phase-dependent variations, but they did not discuss its behavior
in detail. 
Figure~\ref{o5phas} shows the equivalent width indices from fluxes binned over 
the ranges from $-$300 to $-$100~{\kms} and from $-$100 to $+$100~{\kms} 
in the line profile. 
These curves demonstrate that ``blue" and ``central-red" indices follow the 
morphology of the variations in the \ion{C}{4} resonance lines 
(Fig.~\ref{c4n5phas}); i.e., the two halves of the line profiles show 
anticorrelated behavior. 
Thus, the blue half exhibits excess absorption at $\phi$ = 0.5, 
probably because of the importance of the accelerating wind seen
at this orientation. 
In contrast, the red half shows either less absorption or more
emission at this phase. 

We have repeated these extractions for the excited \ion{N}{4}~$\lambda$1718 
line.
Although the curves are noisier, they show the same anticorrelated 
blue/red behavior.

\subsection{Interpreting the red variations}\label{expln}

We now return to the more complicated issue of the origin of the red-wing 
variations, e.g., in the \ion{C}{4} and \ion{N}{5} resonance lines. 
Within the limits imposed by the data quality, these and other optical 
lines listed in Table\,1 show the same variations in the phase range 
0.3\,--\,0.7. This fact suggests that the fluctuations are formed in
the same geometrical structures in the wind.
Because the behavior of this ensemble of lines is similar regardless of
their excitation and ionization potentials, we expect that they are formed 
in a region of thermal equilibrium that is denser than an O-star wind. 
The variations are probably optically thick, though further modeling is
required to demonstrate this explicitly.

There are two basic interpretations for the red-wing variations: either 
they represent {\em excess absorption} at $\phi$ $\approx$
0.9\,--\,0.0 or they represent {\em additional emission} at $\phi$ $\approx$ 
0.5\,--\,0.6.
The distribution of the circumstellar material is implicit
in these interpretations.  
In particular, if the red-wind variations are interpreted as excess
emission, the material must be centered (or nearly centered) on either the
star's magnetic pole or its magnetic equator in order to produce 
equivalent-width curves that are symmetric about the pole-on and equator-on 
phases.
We have identified three scenarios to address this issue:
\begin{description}
\item[{\it Option i:}]
   Central and red-wing {\em absorption} observed at $\phi$ = 0.0 in closed 
   loops projected against the limb of the star. The absorbing wind streams 
   have a net positive velocity component relative to the observer.
\item[{\it Option ii:}]
   Central and red-wing {\em emission} seen in the plane of the sky due to 
   columns of material emanating from the magnetic poles. At
   $\phi$ = 0.5 these columns have rotated to opposite limbs of the star. 
\item[{\it Option iii:}]
   Central and red-wing {\em emission} from infalling material viewed edge-on 
   in the magnetic equatorial plane.
\end{description}

{\it Option i\/} suggests that the variations are due to absorption that is
redshifted with respect to an observer situated above the magnetic pole 
in Fig.\,\ref{cartoon}.
Thus, the gas responsible for it must be moving with a vector component 
directed away from the observer, even though the wind is predominantly 
flowing outward along the magnetic field lines in this region. 
This configuration is difficult to achieve, though it might arise from wind
streams moving along closed loops near the star.
In this case, the line absorption would be visible at phases when the flow 
appears to cross the edge of the stellar disk. 
To explore this idea, we can estimate the fraction of the stellar disk occulted
by such a wind stream in the most ideal circumstances. 
We have computed plasma absorptions using the Hubeny atmosphere and spectral 
synthesis and ``cloud" transfer codes {\it SYNSPEC} and {\it CIRCUS} 
(see SG01 and Smith 2001) for a wind having a temperature of 
35,000\,--\,40,000\,K. 
We find that in the ``best" possible case for {\it Option i,} (i.e., for 
an optically thick wind with no background stellar-limb darkening), an 
absorption component formed at {$-$200~\kms} would have to be produced by 
streams covering at least 9\% of the stellar disk with a column density in H 
of at least 10$^{20}$~cm$^{-2}$ in order to produce the redshifted
variations observed in \ion{C}{4} and \ion{N}{5}. 

  Even so, it is difficult to cover this much of the stellar disk with 
material that is redshifted by as much as $+400$~\kms; see Fig.\,\ref{c4mnmx}.
Since the wind streams near the stellar limb are moving nearly parallel 
to the stellar surface, speeds of {$\sim$1500~\kms} are required to 
account for a projected component of {400~\kms}.
However, the terminal velocity of the wind in this star is no more than 
{2000~\kms} (Walborn \& Nichols 1994): for a line-driven stellar wind 
characterized by a $\beta$=1 velocity law, a velocity as large as 
{1500~\kms} would only be reached at distances of {$\sim$4\,\rstar}; 
i.e., far from the stellar surface.
In view of the implausible degree of fine-tuning required by {\it Option~i},
we do not think it is likely.

{\it Option ii\/} has the advantage that Doppler-shifted emission is
known to be present in some H and He lines at other phases, particularly
$\phi$ = 0, where it extends {$\sim$300~\kms} redward of line center
(Stahl et al. 1993; S96; G. Wade, private communication).
Since the behavior of the UV resonance lines is similar, it is reasonable
to conclude that their variations are also due to changes in emission.

According to {\it Option ii,} scattering brings flux into our line of 
sight when the north magnetic pole rotates to the stellar limb. 
An observer to the right in Fig.\,\ref{cartoon} observes a glowing 
column of particles moving through the plane of the sky. 
However, as the column (``jet") fans out from the plane of the sky,
we expect to see primarily blueshifted H$\alpha$ emission unless the line 
is optically thin. 
Thus, this picture fails to explain the exclusively redshifted emission
observed in \ion{C}{4} and \ion{N}{5}.
Perhaps the idea could be saved by invoking a significant departure from 
magnetic axi-symmetry, e.g., with a decentered dipole and a jet that is 
compressed in one {180\degr} azimuthal sector. 
In such a scheme the observer might see more gas receding over a large area 
of the sky than gas exhibiting blueshifts in the foreground. 
However, even this geometry would produce significant blueshifted 
emission, which is not indicated in Figs.\,\ref{grysc}, \ref{c4mnmx} 
or \ref{n5mnmx}.
Thus, {\it Option ii} is probably not viable.

{\it Option iii} associates the emission maximum at $\phi$ = 0.5
with infalling material near the magnetic equator rather than the poles.
Since the lines of sight to this emission are also those 
that intersect the  high-velocity wind moving toward the observer 
at this phase, this idea seems at first counterintuitive.
How can the wind along the central line of sight to the star 
be approaching and receding at the same time?
Furthermore, this scenario must explain the appearance of redshifted 
in many strong lines, including H$\alpha$ (Stahl et al. 1993). 
In the resonance lines, the emission is probably optically thick. Moreover,
since both low- and high-excitation lines show emission, it is most likely 
formed in a dense circumstellar medium and thus probably close to the star. 
The geometry illustrated in Fig.\,\ref{cartoon} addresses these
issues, in particular by showing how rapidly moving, red- and blue-shifted
material can be found between between the star and a distant observer at 
$\phi$ = 0.5.

\subsection{The infalling wind picture}\label{infall}

The presence of the infalling material required by {\it Option iii} is 
is largely confirmed by recent magnetohydrodynamic (MHD) simulations by 
Dr. A. ud-Doula (private communication; see Gagn\'e et al. 2005 for a 
description of this work), which also provide a physical basis for the 
geometry sketched in Fig.\,\ref{cartoon}. 
As with previous studies of Bp stars, the circulation 
results from the collision of wind streams that originate in opposite 
magnetic hemispheres and are guided by closed loops toward the magnetic 
equator.
In the case of {\thoric}, which has a much denser wind than a typical
Bp star, the collisions at the magnetic equator result in the formation
of dense, optically thick blobs.
These structures cannot be supported by radiation pressure from the star, 
and fall back toward the surface.
Since the blobs are still constrained to move along magnetic fields
lines, they fall obliquely back toward the star along the lines that
originally carried them to the magnetic equator.
Observers viewing the system edge-on ($\phi$ = 0.5) will see the
infall as a redshifted, optically thick feature; in spectra with 
sufficiently high signal-to-noise ratio, this emission might be
variable from one cycle to the next. Ud-Doula's models also include
effects of radiative cooling. His models suggest that the condensations 
are part of the shocks. These structures can be expected to contribute 
to line emission observed in the UV and optical regions. 

Since {\thoric} rotates so slowly, it is appropriate to think of this 
circulation as a rigid, axisymmetric structure comprised of dense blobs.  
We expect that the circulation we have described is related to the 
H$\alpha$ flux and radial-velocity variations (S96; Gagn\'e et al. 2005).  
An additional consistency check for this infall picture is the {\em absence\/}
of evidence for a cool, static disk in the magnetic equatorial plane.
Such disks are the hallmarks of UV-variable Bp stars.
Our modeling of disks for early and late-type Bp stars using {\it SYNSPEC} 
and {\it CIRCUS} (e.g., SG01; Smith 2003) suggests that the disk 
temperature is typically 0.6\,\teff, which for {\thoric} implies a
disk temperature near 25,000\,K.
The ultraviolet line opacity tends to decrease with temperature in this 
temperature range.
Yet even for temperatures as high as 30,000\,K, our modeling experiments 
suggest that a geometrically thin disk with a column density 
of {10$^{22}$ cm$^{-2}$} in H should be detectable as an absorption feature 
near rest velocity in the \ion{C}{4} resonance doublet.

The fact that a static disk is {\em not\/} visible implies that the 
circulating component of the wind interior to the Alfven radius must be 
transported somewhere else. 
We believe the redshifted emission at $\phi$=0.5 is evidence of this 
transport. 
Donati et al. (2002) noted that they were unable to duplicate the 
significantly blueshifted ($\approx -75$~\kms) H$\alpha$ emission in their 
simulation at the pole-on phase ($\phi$ = 0.0). 
They found that the profile was formed mainly from a static disk, which in 
their model has a radius of $\sim$3\,\rstar. 
It is not surprising that the contribution from such a large disk  
would dominate the contribution from the polar wind, but the disk cannot
produce a blueshifted feature at $\phi$ = 0 in this geometry. 

Perhaps the blueshifts are caused by the acceleration of jet-like 
wind from the pole? 
We have investigated this possibility with {\it CIRCUS.} 
We find that even with generous values for relevant parameters the 
contribution of a jet produces negligible H$\alpha$ emission relative to 
the photospheric flux. 
Moreover, it is even more difficult for a jet-emission model to work 
for \ion{He}{2}~$\lambda$4686.
In contrast, our same modeling experiments demonstrate that for similar 
temperatures and densities, a much smaller volume viewed beyond the projected
face of the star could easily produce the observed H$\alpha$ emission. 
From these results we speculate that these blueshifts at $\phi$ = 0 have 
the same cause as the UV resonance line 
redshifts at $\phi$ = 0.5: i.e., they arise from 
the projection of the component of post-shock matter falling obliquely 
toward the star, as indicated by the closed arrowheads in Fig.\,\ref{cartoon}.

\section{Summary}
 
From archival {\iue} spectra of the magnetic O-type star {\thoric}, we 
have discovered new variations of the \ion{C}{4} and \ion{N}{4} resonance 
doublets, as well as the strong, excited lines of \ion{O}{5}~$\lambda$1371 
and \ion{N}{4}~$\lambda$1718.
We found strong similarities between these variations and 
those exhibited by several prominent optical lines reported in the 
literature, including H$\alpha$ and \ion{He}{2}~$\lambda$4686.  
Lines with the largest variations exhibit an anticorrelation in
blue- and red-wing fluxes at $\phi \approx 0.5$: as the red wing 
weakens, the blue wing strengthens.
In addition to this dichotomy, extractions of the profiles over different 
velocity bins show that the ``M"-shaped wave form present in optical lines 
is also present in the \ion{C}{4} lines. 
This morphology disappears at the highest wind velocities sampled (i.e.,
at the most blueshifted wavelengths), where a single-peaked absorption
at $\phi$ = 0.5 replaces the ``M"-pattern.  

The high-velocity strengthenings of the blue wing of \ion{C}{4} at 
$\phi$ = 0.5 are explained by the flow geometry of ud-Doula's MHD simulations. 
These predict that radiative forces open the field lines and carry them 
away from the star once they extend beyond the Alfven radius (i.e., several 
tenths of a stellar radius above the surface; Gagn\'e et al. 2005).  
At $\phi$ = 0.5, the enhanced blue wings of these lines represent
material accelerating away from the star along the magnetic plane.  
We also argued that at $\phi \approx 0.3$ and 0.7 the observer looks along
a smaller projected path length through flows that have moderate outward
velocities. 
This means that the excess absorptions at these phases are weaker than 
the absorptions formed over the longer path length of the open, 
high-velocity flow viewed at $\phi$ = 0.5 (Fig.\,\ref{cartoon}). 

The red variations at the edge-on phases ($\phi = 0.5$), which we interpret 
as emission, require a more complicated geometry in which post-shock material
flows towards the star, likely as dense clumps.
The presence of this material is also predicted by ud-Doula's simulations,
though in the current generation of models the infalling material
only occurs episodically.

Finally, {\it in contrast to bona fide Bp stars, there is no evidence 
for a cool, static disk.} If post-shock material were to accumulate, 
given reasonable extrapolations from models of Bp-star disks,
a disk ought to be visible in spectra of {\thoric}.
We conclude that such accumulation does not happen in the circumstellar
environment of {\thoric}.

\acknowledgements
The authors are grateful to Drs. Marc Gagn\'e, Richard Townsend, Asif
ud-Doula, and Stan Owocki for discussions of their MHD simulations which
have led to a clearer description of our geometry.
We also thank both Dr. Detlef Groote and the referee for their careful 
readings of this paper and helpful comments to improve its clarity. 
Finally, we thank Dr. Gregg Wade for permission to refer to his new
optical data in advance of publication. This work was supported in
part by NASA Grant \#NNG04GE75G.

\clearpage

\centerline {\bf Figure Captions}

\begin{description}

\item[Figure~\ref{cartoon}]
A sketch (not to scale) of the suggested wind/dipolar magnetic-field 
orientation for {\thoric} at $\phi$ = 0.0 (from top), 0.3 and 0.7 
(from 2 o'clock position), and 0.5 (from right), 
assuming angles $i$ = $\beta$ = 45\degr. 
In this planar projection we have collapsed the observer's positions at 
$\phi$ = 0.3 and 0.7, since they are at equal angles in and out of page, 
respectively. 
The stellar wind expands freely only from the regions near the 
magnetic poles (top and bottom of the star).
The magnetic field causes radiatively-driven material from other regions 
of the stellar surface to follow flux lines toward the magnetic equator. 
The wind is divided into two zones according to its position relative to 
the star's Alfven radius, which is designated by the dotted locus and the 
``X" on the line-of-sight from the center of the star to the observer 
(dashed line).
Beyond this radius radiative forces dominate over magnetic forces and cause 
the wind to accelerate away from the star, approximately parallel 
to the magnetic equatorial plane. 
Within the Alfven radius, wind particles emerge from near the magnetic poles, 
follow closed loop lines (open arrowheads), and encounter streams from the 
opposite hemisphere at the magnetic equator, where they shock and accumulate.
Periodically, the particles formerly in the wind condense into optically
thick blobs, which return to the star along the original field lines 
(full arrowheads). 
The redshifted variations suggest that the post-shock gas falls at speeds of 
300\,--\,400~\kms. 

\item[Figure~\ref{grysc}: ]
Dynamic spectrum of the 22 high-dispersion {\iue} echellograms of
{\thoric} obtained through the large-aperture with the SWP camera
showing variations in the \ion{C}{4}~$\lambda\lambda$1548, 1551 doublet 
as a function of magnetic/rotational phase.
The shading represents differences with respect to the time-averaged
mean spectrum (lower panel): darker shades denote phases and wavelengths
where the profile is deeper (i.e., less flux) than its time-averaged value.
The difference spectra have been smoothed over {$\sim$1~\AA}.
Phase gaps in this spectrum are due to the paucity of observations. 
Brackets in the lower panel indicate the extent of the blue and red
components of the doublet for an assumed terminal velocity of 2500~{\kms}
(S96). Although the dominant variations occur in the blue wing of the 
absorption trough (window ``a"), significant variations occur throughout
the profile.  The variations in window ``e" are anti-correlated with 
those in window ``a."

\item[Figure~\ref{c4mnmx}: ]
Three high-resolution {\iue} spectra in the region of the 
\ion{C}{4} resonance lines for phases $\phi$ = 0.51, 0.60, and 0.94 
(SWP\,54040, SWP\,07481, and SWP\,14665).
The units for the flux are erg\,s$^{-1}$\,cm$^{-2}$\,\AA$^{-1}$.
All spectra are binned over 4 pixels, which undersamples the line
cores and give them a sharp appearance. 
The spectra obtained at $\phi$ = 0.60 and 0.94 (solid and dashed lines, 
respectively) exhibit the maximum difference in blue- and red-wing flux.
Wavelength windows {\it a--e} are referenced in Figs.\,\ref{grysc} and 
\ref{c4wng}. 
Following S96, phase zero is defined by the epoch HJD=2448833.0 
(MJD=48832.5).

\item[Figure~\ref{n5mnmx}: ]
{\iue} spectra for phases 0.60 and 0.64 (SWP\,07481 and 54058; solid line is 
their mean) and 0.90 and 0.94 (SWP\,54094 and 14665; dashed line
is their mean) are compared in the region of the \ion{N}{5} doublet of 
{\thoric}. 
The spectra have been binned over 4 points.
The computed difference spectrum, binned over 8 points, is shown at the 
bottom in units of erg\,s$^{-1}$\,cm$^{-2}$\,\AA$^{-1}$\,. 

\item[Figure~\ref{c4n5phas}: ]
Relative equivalent width index (normalized to the maximum value) 
versus phase for the red members of the \ion{C}{4} and and \ion{N}{5}
doublets, as extracted over the velocity range between 0 and $+$250~\kms
(0\,--\,1.3\AA). 
The dotted line is a reference sinusoid.

 \item[Figure~\ref{c4wng}: ]
Equivalent width index versus phase extracted from four velocity ranges 
in the blue wing of \ion{C}{4}~$\lambda$1548; see windows {\it ``a\,--\,e"} 
in Fig.~\ref{grysc}.
The curves have been offset vertically for clarity.
 The corresponding extraction from 1548~\AA\ was not used due to 
interference with the wind absorption of the 1550~\AA\ line.
Note the progression from no variation, to an ``M"-like pattern, 
and finally to a morphology featuring a single peak at $\phi$ $\approx$ 0.5.

\item[Figure~\ref{o5phas}: ]
Equivalent width index variations with phase of the excited 
\ion{O}{5}~$\lambda$1371 line, as extracted from negative and positive 
velocities of the line profile. 

\end{description} 

\begin{figure}
\vspace*{.1in}
\caption{cartoon}
\label{cartoon}
\end{figure}

\begin{figure}
\vspace*{.1in}
\caption{grysc}
\label{grysc}
\end{figure}

\begin{figure}
\vspace*{.1in}
\caption{c4mnmx}
\label{c4mnmx}
\end{figure}

\begin{figure}
\vspace*{.1in}
\caption{n5mnmx}
\label{n5mnmx}
\end{figure}

\begin{figure}
\vspace*{.1in}
\caption{c4n5phas}
\label{c4n5phas}
\end{figure}

\begin{figure}
\vspace*{.1in}
\caption{c4wng}
\label{c4wng}
\end{figure}

\begin{figure}
\vspace*{.1in}
\caption{o5phas}
\label{o5phas}
\end{figure}

\newpage

\begin{deluxetable}{lccccl}
\tablecaption{Dependence of line strength modulations on rotational phase}
\tablecolumns{6}
\tablewidth{6.0 in}
\tablehead{ \colhead{Ion}                            &
            \colhead{$\lambda$ (\AA)}                &
            \colhead{$\chi_{exc}$\tablenotemark{a}}  & 
            \colhead{I.P.\tablenotemark{b}}          &
            \colhead{$\phi$(Max. EW)}                &
            \colhead{Source\tablenotemark{c}}        }
\startdata
\ion{H}{1}  & 6563           & 10.2 &  13.2 & 0.4\,--\,0.6 & S96-4               \\   
\ion{He}{2} & 4686           & 48.4 &  54.4 & 0.4\,--\,0.6 & S96-5               \\
\ion{He}{1} & 4471           & 21.0 &  24.6 & 0.0          & S96-13              \\
\ion{He}{2} & 4541           & 51.0 &  54.4 & 0.0          & S96-14              \\
\ion{C}{4}  & 5801           & 37.5 &  64.5 & 0.0          & S96-15              \\
\ion{C}{4}  & 5812           & 37.5 &  64.5 & 0.0          & R00-7               \\
\ion{C}{4}  & 1548, 1550     &  0.0 &  64.5 & 0.4\,--\,0.6 & S96-8               \\
\ion{C}{4}  & 1548, 1550\tnd &  0.0 &  64.5 & 0.0          & Fig.~\ref{c4n5phas} \\
\ion{N}{4}  & 1718\tnd       & 16.2 &  77.4 & 0.0          & \S~\ref{uvexc}      \\
\ion{N}{5}  & 1238, 1242\tnd &  0.0 &  97.9 & 0.0          & Fig.~\ref{c4n5phas} \\
\ion{O}{3}  & 5592           & 33.9 &  55.0 & 0.0          & S96-16              \\
\ion{O}{5}  & 1371\tnd       & 19.7 & 113.9 & 0.0          & Fig.~\ref{o5phas}   \\
\ion{Si}{4} & 1394, 1402     &  0.0 &  45.1 & 0.4\,--\,0.6 & S96-9               \\
\enddata
\tablenotetext{a}{Excitation potential for the transition in eV.}
\tablenotetext{b}{Ionization potential for the species in eV.}
\tablenotetext{c}{The format ``Ref-n" refers to figure ``n" in reference ``Ref." }
\tablenotetext{d}{Red half only.}

\end{deluxetable}

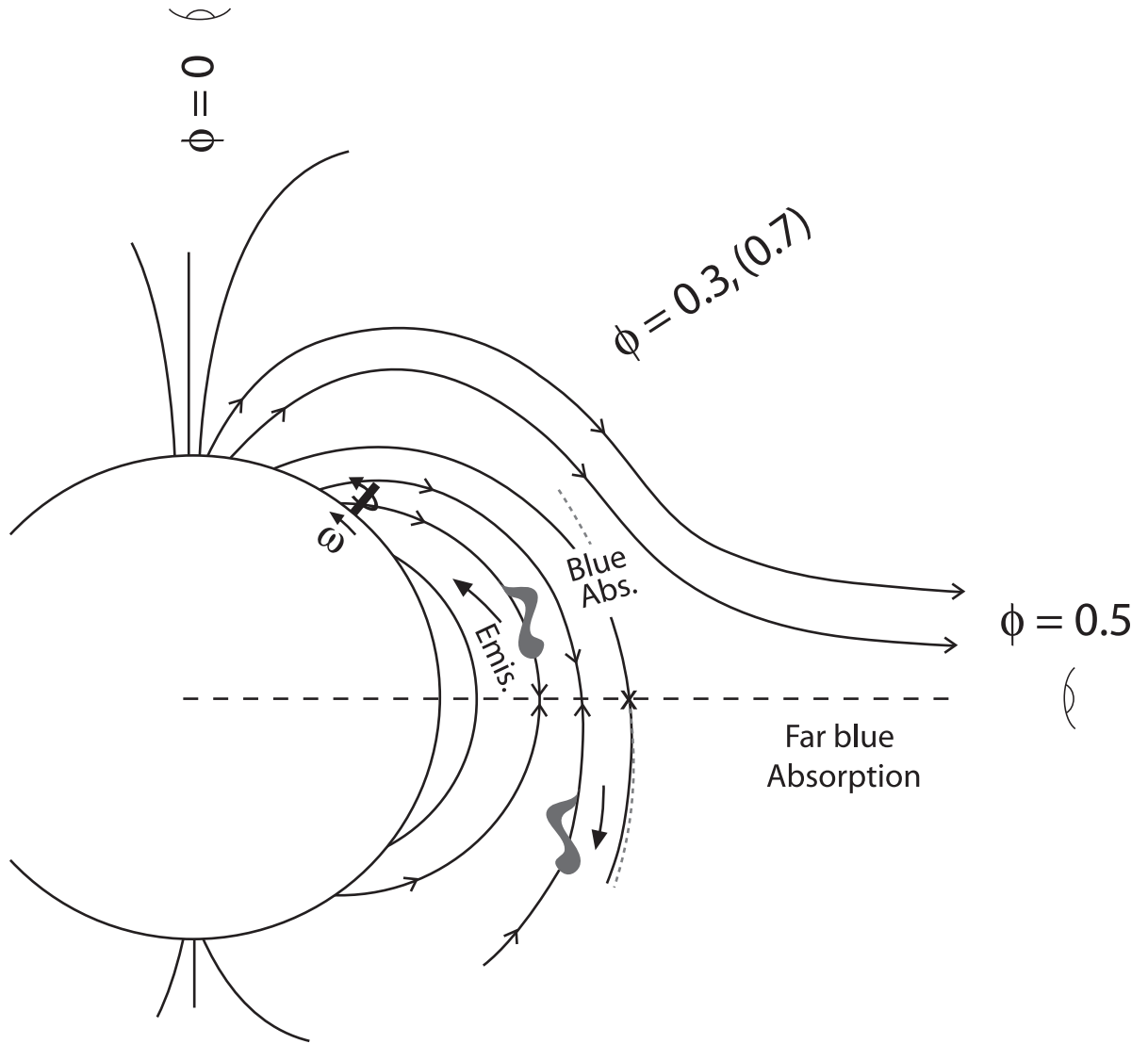
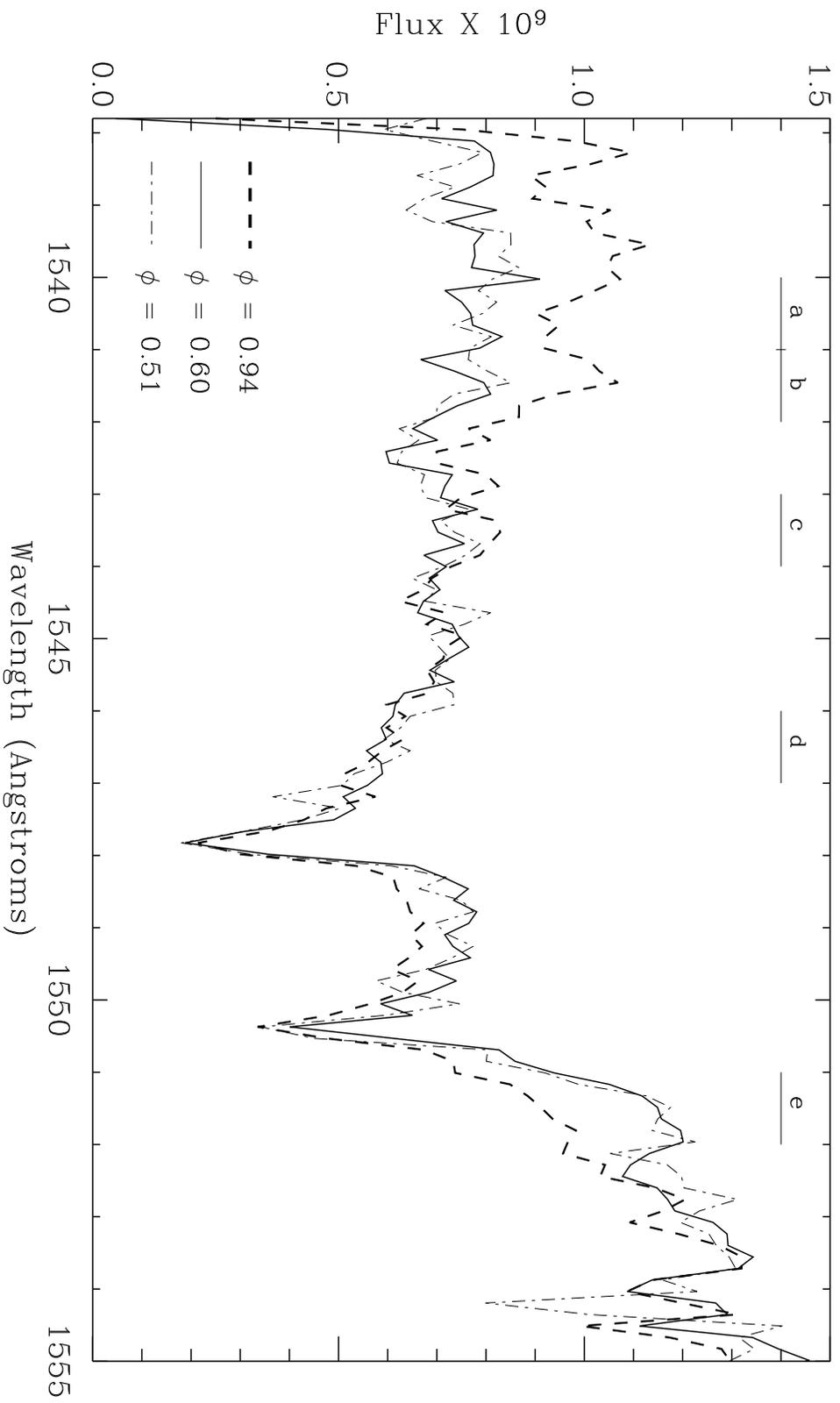
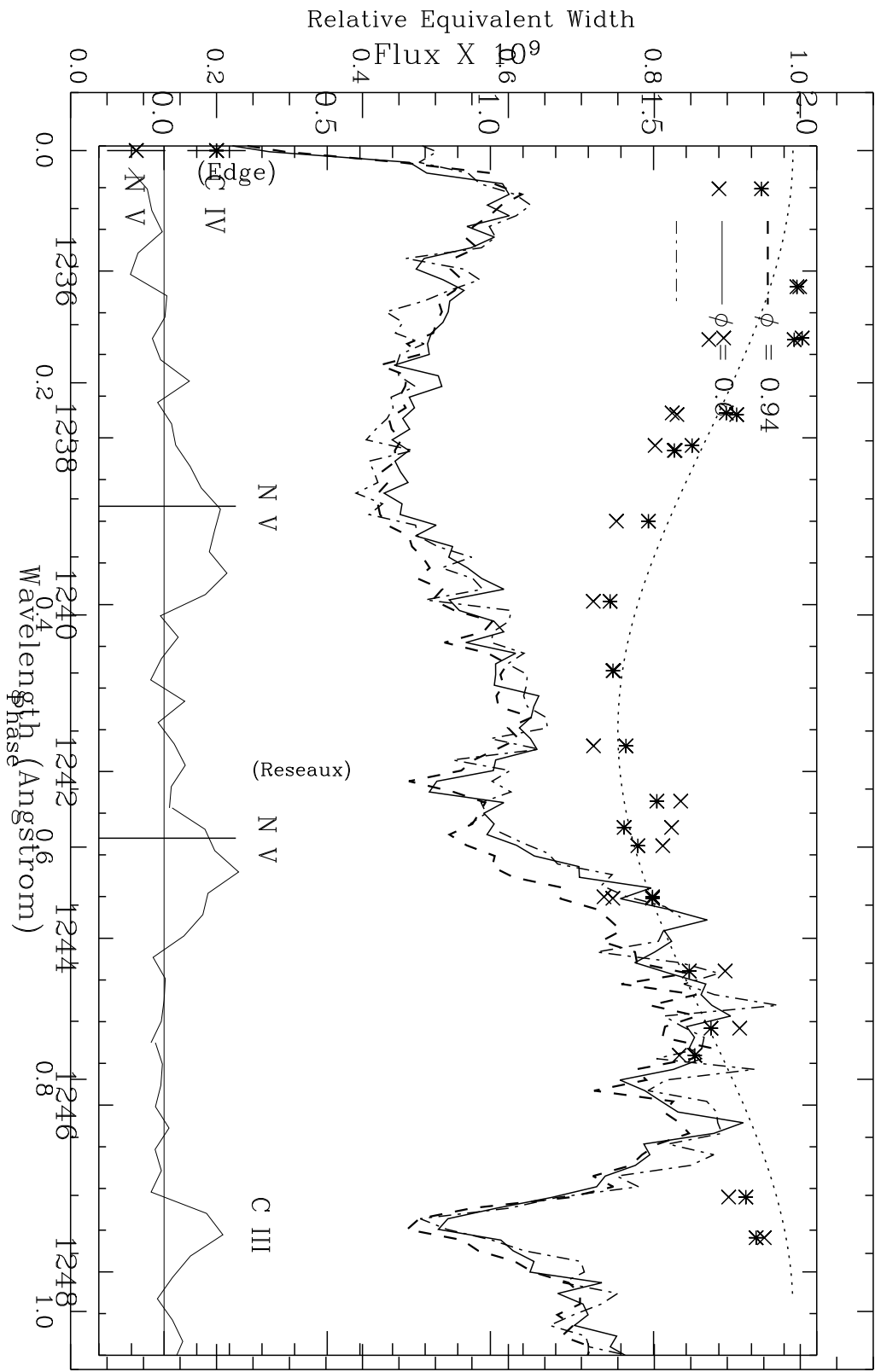
\end{document}